\documentclass{article}
\usepackage{spconf,amsmath,graphicx}

\usepackage{enumitem}
\setlist{nosep, leftmargin=14pt}

\usepackage{mwe} 

\title{Ultrasound Diagnosis of COVID-19: Robustness and Explainability}
\name{
Jay Roberts, Theodoros Tsiligkaridis 
}
\address{  Homeland Sensors and Analytics Group, AI Technology Group \\
  MIT Lincoln Laboratory\\
  Lexington, MA 02421 \\
  \texttt{\{jay.roberts, ttsili\}@ll.mit.edu} \\}
\usepackage{xcolor}
\usepackage{natbib}
\usepackage{graphicx}
\usepackage{amsmath,amssymb,mathtools,amsthm,bm,etoolbox}
\usepackage{subfig}

\newcommand{\EE}{\mathbb{E}}
\newcommand{\loss}{\mathcal{L}}

\DeclareMathOperator*{\argmin}{argmin}
\DeclareMathOperator*{\argmax}{argmax}

\begin{document}
%
\maketitle
\begin{abstract}
Diagnosis of COVID-19 at point of care is vital to the containment of the global pandemic. Point of care ultrasound (POCUS) provides rapid imagery of lungs to detect COVID-19 in patients in a repeatable and cost effective way. Previous work has used public datasets of POCUS videos to train an AI model for diagnosis that obtains high sensitivity. Due to the high stakes application we propose the use of robust and explainable techniques. We demonstrate experimentally that robust models have more stable predictions and offer improved interpretability. A framework of contrastive explanations based on adversarial perturbations is used to explain model predictions that aligns with human visual perception.
\end{abstract}

\section{Introduction}

The Coronavirus disease 2019 (COVID-19) pandemic is the pre-eminent global health crisis of our time. 
Reverse Transcribed Real-Time PCR (RT-PCR) - a molecular test - is one of the most common and effective detection techniques, though there are concerns about processing time and sensitivity of these tests \citep{chen2020recurrence, kanne:2020}. Incorporating medical imaging can be a powerful tool in the diagnostic process. In this paper we consider point of Care Ultrasound (POCUS) which has been shown to be a cost effective and sensitive diagnostic tool \citep{buonsenso2020covid}. Born et al. \cite{born2020pocovidnet} provide an open source database of POCUS imagery and use deep convolutional neural networks (CNN) for automated diagnosis of COVID-19 and bacterial pneumonia \footnote{The dataset and models available at https://github.com/jannisborn/covid19\_pocus\_ultrasound.}. While it has been shown that high diagnostic accuracy is achievable using deep CNNs for frame- and video-based detectors \citep{Born:Exp:2020}, the issue of model robustness in this context has not received attention.

\begin{figure}[h]
\centering
  \subfloat[Standard CNN]{
    \includegraphics[height=0.11 \textwidth]{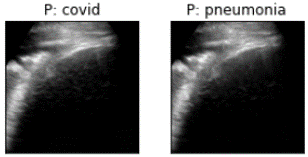}
    }\label{fig:vgg st comp}
\subfloat[Robust CNN]{
    \includegraphics[height=0.11 \textwidth]{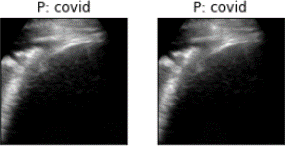}
    }\label{fig:vgg comp comp}
    \caption{
    Examples of $L^2$ Adversarial perturbations against a standard (a) and robust (b) ResNet18 network attempting to fool the models correct prediction (P) of COVID. The input image is shown on the left and the perturbed image on the right.
    }
\label{fig:vgg pert}
\end{figure}

It has been well established that large capacity deep learning models can be fooled by small, carefully chosen, input perturbations known as adversarial attacks \citep{goodfellow:2015}. In the context of computer vision such attacks can be imperceptible to the human eye but still manage to fool the model. Though we do not imagine that medical diagnosis systems will be subjected to such attacks from malicious actors, their existence implies that the model may be learning features which are not medically relevant. This degrades trust from medical practitioners and brings into question the validity of any proposed explanation of the model's predictions. 
Figure \ref{fig:vgg pert} (a) and (b) shows examples of such perturbations. 

The adoption of AI-enabled techniques for augmenting diagnostic decision processes used by clinicians depends on the important issue of trust. Both robustness to imperceptible changes and explainability for justifying model predictions are critical factors in establishing trust. Our preliminary work presented here paves the way for providing insight into how robust deep CNNs make diagnostic decisions using ultrasound imaging modality by visualizing feature changes that lead to different decision boundaries, in addition to providing robustness.

\section{Methods}



The prevalent way of training neural networks is through the empirical risk minimization (ERM) principle defined as 
\begin{equation}
\label{eq:std train}
    \min_\theta \EE_{(x,y)\sim \mathcal{D}} \left[\ 
                    \loss(x ,y;\theta)
                    \ 
                    \right].
\end{equation}
This yields high accuracy on test sets but leaves the network vulnerable to adversarial attacks. 
An effective defense against such attacks is adversarial training (AT) \cite{madry2018:at} which instead aims to minimize the adversarial risk 
\begin{equation} 
\label{eq:adv_risk}
    \min_\theta \EE_{(x,y)\sim \mathcal{D}} 
        \left[ 
            \max_{\delta \in B_p(\epsilon)}  
            \loss(x+\delta,y;\theta) \right]
            .
\end{equation}
The training procedure constructs adversarial attacks, $\delta$, at given inputs, $x$, that aim to maximize the loss, $\loss$. The attacks are constrained to be within some $\epsilon$, in the $L^p$ sense, of the original image, ensuring that the perturbed image resembles the original. Common choices of constraint are $L^2$ and $L^\infty$. Though $L^2$ perturbations can be more noticeable than  $L^\infty$ perturbations, they are generally smoother and so for the purposes of visual explanations we restrict our robustness discussion to the $L^2$ case.

A common method to approximate the maximization is projected gradient descent (PGD) which performs iterative updates of the approximation based on the gradient of the loss. 
Because of the high number of forward-backward propagations AT can become computationally expensive for large and/or high resolution datasets. There are several regularization methods aimed at obtaining robustness with less computation \citep{Lyu:2015, md2019:cure, scorpio}. However, since the POCUS dataset is small we use AT to train our robust models.

\subsection{Explanations}

As noted in \citep{ilyas2019adversarial}, adversarial attacks are not strictly a detriment and can be used to discern concepts that a model has learned. For robust models in particular these features have shown to be better aligned with human perception than their non-robust counterparts. To this end, we consider a framework similar to \citep{scorpio}.
The approach finds \textit{pertinent negatives/positives} by optimizing over the perturbation variable $\delta$. Pertinent negatives capture what is missing in the prediction and pertinent positives refer to critical features that are present in the input examples.

We consider two contrastive
explanations defined by the optimizations
\begin{equation} \label{eq:L2attack}
    \delta_{\max}
        := \argmax_{\delta \in B_2(\epsilon)} l(x+\delta,y)
\end{equation}
and
\begin{equation}
    \delta_{\min}
        :=\argmin_{\delta \in B_2(\epsilon)} l(x+\delta,y)
    ,
\end{equation}
where the losses are locally optimized within an $L^2$ ball of radius $\epsilon$ to ensure that the perturbation $\delta$ is small. 

In the case of correct predictions, $\delta_{\max}$ are features that can be added in
the original image which flip the model's decision to a nearby class and therefore are \emph{pertinent negative features} of the image. 
Whereas $\delta_{\min}$ are features which contribute to the correct prediction, making them the \emph{pertinent positive features}. In the case of incorrect predictions the roles are flipped. The $\delta_{\max}$ are features which can be added or emphasized to make the model more confident in its incorrect prediction; the $\delta_{\min}$ are the features that are missing for the model to make a correct prediction.

This framework can be used to investigate features a model has learned and to identify trends in failure cases. 

\section{Results}

 We used lung point-of-care ultrasound (POCUS) imagery gathered by \citep{born2020pocovidnet} to train our network. 
 This is the first publicly available dataset of lung POCUS recordings of COVID-19, bacterial pneumonia, and healthy patients.
The dataset consists of 3119 frames from 195 ultrasound videos. A 5-fold cross validation over the videos was performed.
 We evaluated two deep CNNs on the POCUS dataset VGG16 \citep{simonyan2015deep} and ResNet18 \citep{he2015deep}.
Models were trained for 51 epochs using an SGD optimizer that had an initial learning rate of $0.01$ which decayed by a factor of 10 every 15 epochs. In this paper, we report results for the best models over the 51 epochs. For standard models (trained with ERM) the best model is defined as the model with highest clean accuracy. For robust models (trained with AT) it is the model with the highest adversarial accuracy.

{\bf Performance.} 
Table \ref{tab:outcome table} shows the performance for each outcome task. All models performed well in pneumonia detection and in all tasks VGG16 outperforms its ResNet18 counterpart. For the remaining outcomes the robust models have less sensitivity than their standard counterparts. 

\begin{figure}[h]
    \centering
    \includegraphics[width=0.45\textwidth]{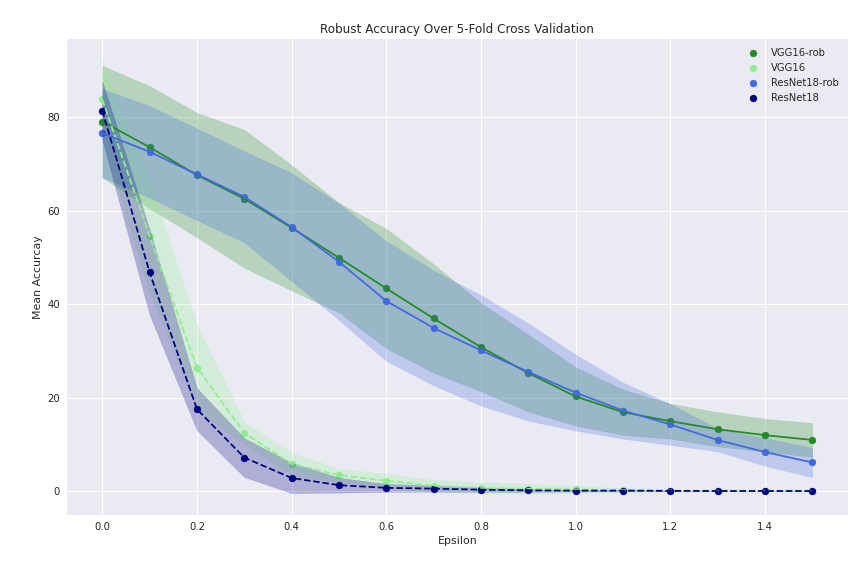}
    \caption{Accuracy of models against increasing strength of L2 attacks. The mean is plotted with the shaded area being one standard deviation across splits. Standard models (dashed) experience dramatic degradation of performance compared to robust models (solid).}
    \label{fig:my_label}
\end{figure}

Figure \ref{fig:my_label} shows the overall performance of robust models versus standard models across the 5-fold cross validation for increasingly strong adversarial attacks. We see the standard models perform well in the clean setting (epsilon = 0), achieving mean accuracy above 80\% and outperforming their robust counterparts. However, the performance of standard models degrades dramatically as the attack strength increases compared to the robust models, which maintain better performance. This suggests that the standard models have learned brittle features sensitive to idiosyncrasies and/or noise in the training dataset. The robust models lacked such sensitivity, suggesting they learned more medically relevant features. Further, as the attack strength increases the robust models eventually lose performance - which is expected since a model that is robust to very large perturbations may be insensitive to small but medically relevant features in the input.

\begin{table}[h]
\begin{tabular}{llrr}
Model &  Outcome &    Acc.     &    AUROC     \\
\hline
ResNet18-rob & covid &  78.076 &  81.326 \\
         & pneumonia &  93.994 &  95.066 \\
         & regular &  81.146 &  81.372 \\
\hline
ResNet18 & covid &  82.616 &  85.746 \\
         & pneumonia &  93.856 &  95.992 \\
         & regular &  86.422 &  87.992 \\
\hline
VGG16-rob & covid &  81.498 &  87.054 \\
         & pneumonia &  93.568 &  96.952 \\
         & regular &  83.122 &  85.114 \\
\hline
VGG16    & covid &  85.992 &  89.646 \\
         & pneumonia &  95.144 &  97.618 \\
         & regular &  86.508 &  88.928 \\
\end{tabular}
\caption{Mean clean accuracy (Acc.) and AUROC over 5 splits for each outcome for VGG16 and ResNet18 trained with standard ERM or with AT (rob).}
    \label{tab:outcome table}
\end{table}

{\bf Explanations.} 

\begin{figure}[h]
\centering
  \subfloat[Robust-Correct-Min]{
    \includegraphics[width=0.22 \textwidth]{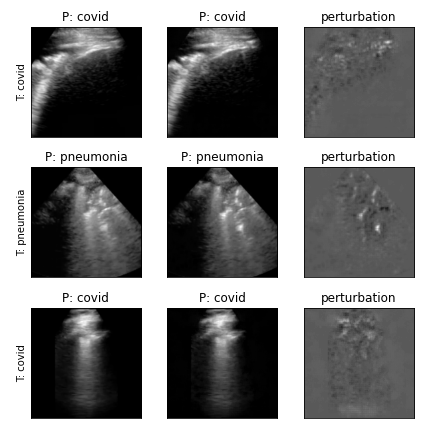}
    }\label{fig:delta min corr rob}
    \hfill
\subfloat[Standard-Correct-Min]{
    \includegraphics[width=0.22 \textwidth ]{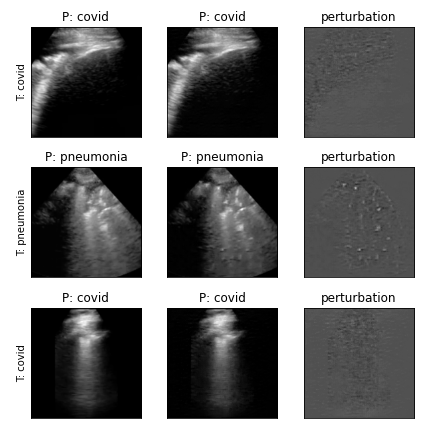}
    }\label{fig:delta min corr std}
    \hfill
      \subfloat[Robust-Error-Min]{
    \includegraphics[width=0.22 \textwidth ]{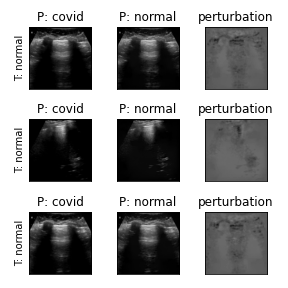}
    }\label{fig:delta min err rob}
    \hfill
\subfloat[Standard-Error-Min]{
    \includegraphics[width=0.22 \textwidth ]{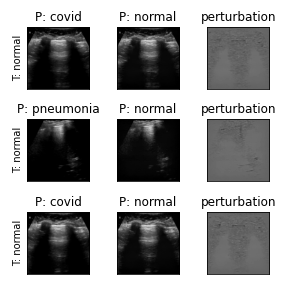}
    }\label{fig:delta min err std}
    \caption{Examples of a $\delta_{min}$ perturbation for robust and standard models. The columns are the input image, adversarial perturbed image, and the perturbation respectively. The target (T) label is given on the left of each row and the model's prediction (P) is given above each image.}
    \label{fig:delta min}
\end{figure}

\begin{figure}[h]
  \subfloat[Robust-Correct-Max]{
    \includegraphics[width=0.22 \textwidth]{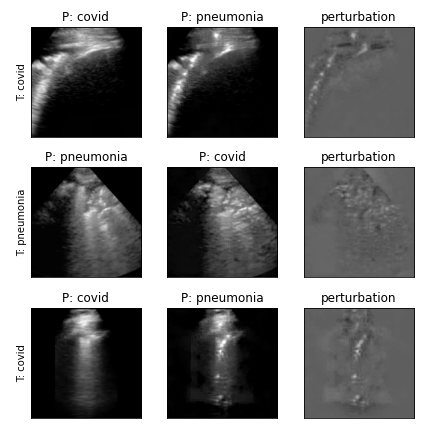}
    }\label{fig:delta max corr rob}
    \hfill
\subfloat[Standard-Correct-Max]{
    \includegraphics[width=0.22 \textwidth]{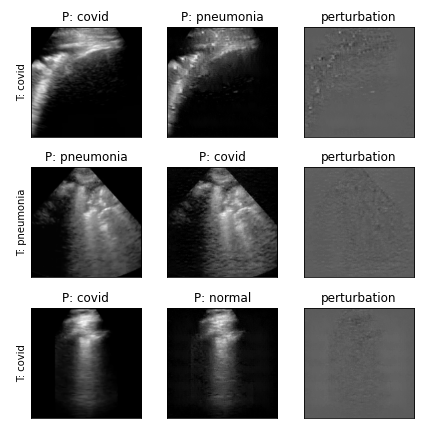}
    }\label{fig:delta max corr std}
    \hfill
      \subfloat[Robust-Error-Max]{
    \includegraphics[width=0.22 \textwidth]{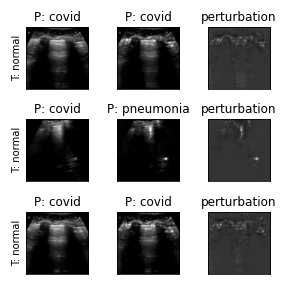}
    }\label{fig:delta max err rob}
    \hfill
\subfloat[Standard-Error-Max]{
    \includegraphics[width=0.22 \textwidth]{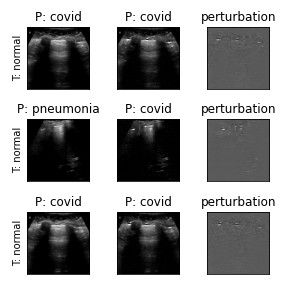}
    }\label{fig:delta max err std}
    \caption{
   Examples of a $\delta_{max}$ perturbation for robust and standard models. The columns are the input image, adversarial perturbed image, and the perturbation respectively. The target (T) label is given on the left of each row and the model's prediction (P) is given above each image. 
    }
\label{fig:delta max}
\end{figure}

Figures \ref{fig:delta min} and \ref{fig:delta max} show $\delta_{\min}$ and $\delta_{max}$ perturbations for robust and standard models. In general, the perturbations of robust models are more focused and targeted than the perturbations for standard models which were diffuse, making their learned features less interpretable. A notable exception is row-2 of Figure \ref{fig:delta min} where both models seem to focus on the pocket-like features present in many bacterial pneumonia POCUS imagery. 

Figure \ref{fig:delta min} shows the effects of a $\delta{\min}$ attack
on robust and standard networks for correct predictions
and errors. For the pertinent positives of the correct predictions
(Figure \ref{fig:delta min} (a) and (b)) both models emphasize
the pockets seen in the pneumonia image but for the other
two images the standard model seems to only be focusing on
the brighter parts of the image. In comparison, the robust model has
picked up on distinct features of the original image. 

The pertinent negatives (Figure \ref{fig:delta min} (c) and (d)) of the robust model appear to de-emphasize certain features of the input. This is particularly interesting since these were images of healthy lungs which the models mistook for pathology. These perturbations allow us to not only identify difficult cases for the models to classify, but to effectively localize the features that fooled the model. Such an insight can be used to identify gaps in training data or flag common failure cases.

Figure \ref{fig:delta max} shows the effects of a $\delta_{max}$ attacks. For the correct predictions (Figure \ref{fig:delta max} (a) and (c)) these are the pertinent negative features that would cause the network to misclassify. The perturbations for robust models suggest they have learned more relevant features than the standard model. For example, in rows 1 and 3 we see that the pocket-like features are missing from the COVID-19 images that would cause the model to classify them as pneumonia. Whereas in row 2, the perturbation has removed some pockets and fused others together. In contrast, the perturbations for the standard model are less interpretable. Figure \ref{fig:delta max} (b) and (d) show the pertinent positive features that led the model to misclassify. It is clear that the robust model focuses on the upper region for its predictions while the standard model used more diffuse features.

\section{Conclusion and Future Work}
In this work we advocate for using robust AI in the safety-critical domain of automated diagnosis. We show that while standard models may outperform robust ones in terms of raw metrics, it comes at the cost of reliability and explainability. Further, we provide a means of using adversarial attacks to discern the features learned by robust models. Future work will include collaborating with hospitals to leverage more ultrasound data - improving accuracy, quality of robust features, and allowing us to focus on discriminating between multiple manifestations of COVID-19 (e.g. B-lines, pleural abnormalities, consolidations). Additionally, we plan to guide our explanation framework based on expert radiologist feedback.



\section{Acknowledgements}

This material is based upon work supported by the Under Secretary of Defense for Research and Engineering under Air Force Contract No. FA8702-15-D-0001. Any opinions, findings, conclusions or recommendations expressed in this material are those of the author(s) and do not necessarily reflect the views of the Under Secretary of Defense for Research and Engineering.

\section{Compliance with Ethical Standards}

This research study was conducted retrospectively using human subject data aggregated, processed, and made available in open access by Born et. al. \cite{born2020pocovidnet} 
. Up to date references of data sources and access can be found at :
https://github.com/jannisborn/covid19\_pocus\_ultrasound. 
Ethical approval was not required as confirmed by the license attached with the open access data.

\bibliographystyle{plainnat}
\bibliography{main}

\end{document}